\documentclass[11pt]{article}
\usepackage{amsmath,epsfig,sint}
\font\sixrm=cmr6

\def\rmd{{\rm d}}

\def\rmO{{\rm O}}

\def\defeq{\mathrel{\mathop=^{\rm def}}}
\def\proof{\noindent{\sl Proof:}\kern0.6em}

\def\frac#1#2{\hbox{$#1\over#2$}}
\def\dual{\mathstrut^*\kern-0.1em}

\def\lvec#1{\setbox0=\hbox{$#1$}
    \setbox1=\hbox{$\scriptstyle\leftarrow$}
    #1\kern-\wd0\smash{
    \raise\ht0\hbox{$\raise1pt\hbox{$\scriptstyle\leftarrow$}$}}
    \kern-\wd1\kern\wd0}
\def\rvec#1{\setbox0=\hbox{$#1$}
    \setbox1=\hbox{$\scriptstyle\rightarrow$}
    #1\kern-\wd0\smash{
    \raise\ht0\hbox{$\raise1pt\hbox{$\scriptstyle\rightarrow$}$}}
    \kern-\wd1\kern\wd0}

\def\nabstar#1{\nabla\kern-0.5pt\smash{\raise 4.5pt\hbox{$\ast$}}
               \kern-4.5pt_{#1}}
\def\drv#1{{\partial_{#1}}}
\def\drvstar#1{\partial\kern-0.5pt\smash{\raise 4.5pt\hbox{$\ast$}}
               \kern-5.0pt_{#1}}

\def\momp#1#2{
    \setbox0=\hbox{${#1}$}\setbox1=\hbox{${#1}_{#2}$}
    {#1}_{#2}\kern-\wd1\kern\wd0
    \smash{\raise4.5pt\hbox{$\scriptscriptstyle +$}}}
\def\momm#1#2{
    \setbox0=\hbox{${#1}$}\setbox1=\hbox{${#1}_{#2}$}
    {#1}_{#2}\kern-\wd1\kern\wd0
    \smash{\raise4.5pt\hbox{$\scriptscriptstyle -$}}}
\def\mompm#1#2{
    \setbox0=\hbox{${#1}$}\setbox1=\hbox{${#1}_{#2}$}
    {#1}_{#2}\kern-\wd1\kern\wd0
    \smash{\raise4.5pt\hbox{$\scriptscriptstyle \pm$}}}
\def\smomp#1#2{
    \setbox0=\hbox{${#1}$}\setbox1=\hbox{${#1}_{#2}$}
    {#1}_{#2}\kern-\wd1\kern\wd0
    \smash{\raise3pt\hbox{$\scriptscriptstyle +$}}}
\def\smomm#1#2{
    \setbox0=\hbox{${#1}$}\setbox1=\hbox{${#1}_{#2}$}
    {#1}_{#2}\kern-\wd1\kern\wd0
    \smash{\raise3pt\hbox{$\scriptscriptstyle -$}}}
\def\smompm#1#2{
    \setbox0=\hbox{${#1}$}\setbox1=\hbox{${#1}_{#2}$}
    {#1}_{#2}\kern-\wd1\kern\wd0
    \smash{\raise3pt\hbox{$\scriptscriptstyle \pm$}}}

\def\si{\kern1pt{\rm si}}
\def\co{\kern1pt{\rm co}}

\def\Nf{N_{\rm f}}
\def\psibar{\overline{\psi}}

\def\rhoprime{\rho\kern1pt'}
\def\rhobar{\bar{\rho}}
\def\rhobarprime{\rhobar\kern1pt'}
\def\rhobartilde{\kern2pt\tilde{\kern-2pt\rhobar}}
\def\rhobartildeprime{\kern2pt\tilde{\kern-2pt\rhobar}\kern1pt'}

\def\zetabar{\bar{\zeta}}
\def\zetaprime{\zeta\kern1pt'}
\def\zetabarprime{\zetabar\kern1pt'}
\def\zetar{\zeta_{\raise-1pt\hbox{\sixrm R}}}
\def\zetabarr{\zetabar_{\raise-1pt\hbox{\sixrm R}}}

\def\phiimpr{\phi_{\kern0.5pt\hbox{\sixrm I}}}

\def\ar{A_{\hbox{\sixrm R}}}
\def\pr{P_{\hbox{\sixrm R}}}

\def\rat{\rho}

\def\dirac#1{\gamma_{#1}}
\def\diracstar#1#2{
    \setbox0=\hbox{$\gamma$}\setbox1=\hbox{$\gamma_{#1}$}
    \gamma_{#1}\kern-\wd1\kern\wd0
    \smash{\raise4.5pt\hbox{$\scriptstyle#2$}}}

\def\ba{b_{\rm A}}
\def\bp{b_{\rm P}}

\def\ca{c_{\rm A}}

\def\csw{c_{\rm sw}}

\def\ct{c_{\rm t}}

\def\ctildet{\tilde{c}_{\rm t}}

\def\fp{f_{\rm P}}

\def\f1{f_1}

\def\h1{h_1}

\def\CF{C_{\rm F}}
\def\cf{\CF}

\def\opprime#1{\setbox0=\hbox{${\cal O}$}\setbox1=\hbox{${\cal O}_{\rm #1}$}
    {\cal O}_{\rm #1}\kern-\wd1\kern\wd0
    \smash{\raise4.5pt\hbox{\kern1pt$\scriptstyle\prime$}}\kern1pt}

\def\ophatprime#1{\setbox0=\hbox{$\widehat{\cal O}$}
    \setbox1=\hbox{$\widehat{\cal O}_{\rm #1}$}
    \widehat{\cal O}_{\rm #1}\kern-\wd1\kern\wd0
    \smash{\raise4.5pt\hbox{\kern1pt$\scriptstyle\prime$}}\kern1pt}

\def\bopprime#1{\setbox0=\hbox{${\cal O}$}\setbox1=\hbox{${\cal O}_{\rm #1}$}
    {\cal L}_{\rm #1}\kern-\wd1\kern\wd0
    \smash{\raise4.5pt\hbox{\kern1pt$\scriptstyle\prime$}}\kern1pt}

\def\blagprime#1{\setbox0=\hbox{${\cal B}$}\setbox1=\hbox{${\cal B}_{#1}$}
    {\cal B}_{#1}\kern-\wd1\kern\wd0
    \smash{\raise5.2pt\hbox{\kern1pt$\scriptstyle\prime$}}\kern1pt}

\def\opprime{{{\cal O}^{\kern1pt\smash{\hbox{$\scriptstyle\prime$}}}}}

\def\gbar{\bar{g}}

\def\gr{g_{{\hbox{\sixrm R}}}}

\def\mq{m_{\rm q}}

\def\mrs{m_{{\hbox{\sixrm R,s}}}}

\def\mc{m_{\rm c}}

\def\d1SF{d_1^{\hbox{\sixrm SF}}}
\def\b2SF{b_2^{\hbox{\sixrm SF}}}

\def\za{Z_{\rm A}}

\def\zp{Z_{\rm P}}

\def\zzeta{Z_{\zeta}}

\def\Zp{\zp}

\def\ms{{\rm MS}}
\def\msbar{{\rm \overline{MS\kern-0.05em}\kern0.05em}}

\def\smallSF{\hbox{\sixrm SF}}

\def\mbar{\kern1pt\overline{\kern-1pt m\kern-1pt}\kern1pt}

\newcommand{\bes}{\begin{eqnarray}}
\newcommand{\ees}{\end{eqnarray}}

\newcommand{\fraction}[2]{{#1\over#2}}

\begin{document}
\begin{titlepage}
\begin{flushright}
   MPI-PhT/98-50\\
   FSU-SCRI-98-74\\
   August 1998
\end{flushright}

\vskip 1 cm
\begin{center}
  {\Large\bf  The running quark mass in the SF scheme\\[1.5ex]
   and its two-loop anomalous dimension}
\end{center}
\vskip 1 cm
\begin{figure}[h]
\begin{center}
\epsfig{figure=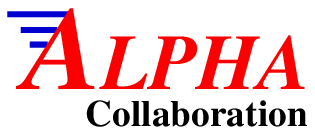} 
\end{center}
\end{figure}
%\vskip 1 cm
\begin{center}
{\large Stefan Sint$^{\scriptscriptstyle a}$
    and Peter Weisz$^{\scriptscriptstyle b}$}
\vskip 2.3ex
$^{\scriptstyle a}$ SCRI, Florida State University\\
Tallahassee, Florida 32306--4130\\
\vskip 1.5ex
$^{\scriptstyle b}$ Max-Planck-Institut f\"ur Physik\\
F\"ohringer Ring 6, D-80805 M\"unchen, Germany\\
\vskip 1.5cm
{\bf Abstract}
\vskip 0.7ex
\end{center}

The non-perturbatively defined running quark mass introduced by
the ALPHA collaboration is based on the PCAC relation between
correlation functions derived from the Schr\"odinger functional (SF).
In order to complete its definition it remains to specify a number of
parameters, including the ratio of time to spatial extent, $T/L$,
and the angle $\theta$ which appears in the spatial boundary conditions
for the quark fields. We investigate the running mass in perturbation
theory and propose a choice of parameters which attains two
desired properties: firstly the two-loop anomalous dimension
$\d1SF$ is reasonably small. This is needed in order
to ease matching with the non-perturbative computations and
to achieve a precise determination of the renormalization group invariant
quark mass. Secondly, to one-loop order of perturbation theory, cut-off
effects in the step-scaling function are small in O($a$) improved lattice 
QCD.

\vfill

%\begin{flushleft}
%  MPI-PhT/98-50\\
%  FSU-SCRI-98-nn \\
% June 1998
%\end{flushleft} 

\eject

\vfill

\eject

\end{titlepage}

\section{Introduction}

This paper is part of a project by the ALPHA collaboration
to determine the $\Lambda$-parameter 
and the renormalization group invariant quark masses
with controlled errors, using hadronic observables as
experimental input~\cite{letter,Martintalk}.
For numerical simulations of the lattice regularized theory,
the basic difficulty consists in the large difference
of length scales, ranging from the long distances typical for hadronic
physics to short distances where perturbation theory can be applied
with confidence. The proposed solution~\cite{letter} 
combines an intermediate finite volume renormalization scheme
with a finite size scaling technique~\cite{LWW}, which allows to 
step up the energy ladder recursively.

The very definition of the renormalized running coupling
and quark mass is of great importance for the method to be practical.
The running parameters should be relatively easy to compute by
numerical simulation and they should not be affected by large cutoff
effects. It is then possible to perform reliable continuum extrapolations
and trace the non-perturbative evolution of the running parameters
directly in the continuum limit, covering a wide range of energy scales.
At high energies the evolution may be compared with perturbation
theory, and, once perturbative evolution has set in, one
may use perturbation theory to evolve to infinite
energy and determine the renormalization group invariant parameters.
For this last step to be feasible, the matching to
perturbation theory should not require extremely high
energies. This can be regarded as a further requirement to be met by
a sensible definition of the running parameters.

In this paper we carry out a perturbative investigation of
a two-parameter family of running quark masses in the SF
scheme~\cite{letter}, which depends on the ratio $\rho=T/L$ between time and
spatial extent of the space-time manifold, as well as on the parameter
$\theta$ appearing in the spatial boundary conditions 
on the quark fields. In particular we determine 
the two-loop quark mass anomalous dimension $\d1SF$ 
which is needed for a precise determination of the renormalization 
group invariant quark mass~\cite{RainerEtAl}. 
It turns out that the parameters must be chosen 
with care for $\d1SF$ to be reasonably small. Taking
the size of the one-loop cutoff effects in the step-scaling function
as a further criterion suggests a specific choice of both parameters.
This completes the definition of the running quark mass in the
SF scheme which is expected to meet all of the above mentioned requirements.

Many technical details of our perturbative calculation have 
appeared in ref.~\cite{paperV} and will not be repeated here.
Parts of our results as well as of the corresponding 
non-perturbative study~\cite{RainerEtAl}
have already been published in ref.~\cite{Martintalk}.
This paper is organised as follows: In sect.~2 we recall 
some well-known facts about the renormalization group,
in particular the relation between running parameters
and the renormalization group invariants.
We then review the SF scheme and present our perturbative results 
for the renormalized quark mass in the continuum limit (sect.~3). 
Cutoff effects are discussed in sect.~4 and we end with a short summary.
Finally, an appendix has been included to indicate the
changes to appendix~A of ref.~\cite{paperV} and appendix~B of 
ref.~\cite{paperII} for our more general choice of parameters.

\section{Renormalization group}

In order to put the present work into its context 
we  review some aspects of the renormalization 
group for QCD with $N$ colors and $\Nf$ quark flavors
and a diagonal quark mass matrix. 

\subsection{Callan Symanzik equation and running parameters}

In the following it is assumed that the theory
has been regularized, e.g.~through the introduction of a 
space-time lattice, and that all renormalization conditions 
are independent of the quark masses. Any physical quantity 
$P$ is a renormalization group invariant,~i.e.
as a function of the normalization mass 
$\mu$, the renormalized coupling $\gr$ and
the renormalized quark masses $\mrs$, $s=1,\dots,\Nf$, it 
satisfies the Callan-Symanzik equation, 
\begin{equation}
  \left\{ \mu\fraction{\partial}{\partial\mu} 
       +\beta(\gr)\fraction{\partial}{\partial\gr}
       +\tau(\gr)\sum_{s=1}^{\Nf}\mrs\fraction{\partial}{\partial\mrs}
  \right\} P =0.
  \label{CS}
\end{equation}
The renormalization group functions $\beta$ and $\tau$ 
can be derived from the relation between the bare and
renormalized parameters, using the independence of the
former upon $\mu$. The precise formulae depend on the  
details of the regularization and will not be needed in the following.

For small couplings $\beta$ and $\tau$
admit asymptotic expansions of the form
\begin{eqnarray}
  \beta(g) &\buildrel{g}\rightarrow0\over\sim
            & -g^3\sum_{k=0}^\infty b_k g^{2k},\\ 
  \tau(g)  &\buildrel{g}\rightarrow0\over\sim
            & -g^2\sum_{k=0}^\infty d_k g^{2k}, 
\end{eqnarray}
with coefficients which are
renormalization scheme dependent in general.
In the minimal ($\ms$) or modified minimal ($\msbar$) scheme 
of dimensional regularization $b_k$ and $d_k$ 
are known for $k\leq 3$~\cite{betatauI-betatauIII}, 
the first few being given by
(with $\cf=(N^2-1)/2N$),
\begin{eqnarray}
  b_0 &=& \bigl\{\frac{11}{3}N-\frac23\Nf\bigr\}(4\pi)^{-2},\\[1ex]
  d_0 &=&  6\,\cf(4\pi)^{-2},\\[1ex]
  b_1 &=& \bigl\{\frac{34}{3}N^2-(\frac{13}{3}N-N^{-1})\Nf\bigr\}(4\pi)^{-4},
         \\[1ex]
  d_1 &=& \cf\bigl\{\frac{203}{6}N-\frac32 N^{-1}
          -\frac{10}{3} \Nf\bigr\}(4\pi)^{-4}. 
  \label{d1msbar}
\end{eqnarray}
While the minimal schemes of 
dimensional regularization are only defined in a perturbative framework,
we emphasize that $\beta$ and $\tau$ are 
in general non-perturbatively 
defined functions. If these are given, running parameters 
at the momentum scale $q$ are obtained by integrating the equations
\begin{equation}
   q {{\partial\gbar}\over{\partial q}} = \beta(\gbar),\qquad
   q {{\partial\mbar_s}\over{\partial q}} = \tau(\gbar)\mbar_s, 
\end{equation}
with the boundary conditions
\begin{equation}
  \gbar(\mu)=\gr, \qquad
  \mbar_s(\mu)=\mrs, \quad s=1,\dots,\Nf.
\end{equation}
The running parameters are related to the renormalization group
invariant (RGI) quark masses $M_s$ and the $\Lambda$ parameter, through
($s=1,\ldots,\Nf$),
\begin{eqnarray}
  \Lambda &=& q\,(b_0\gbar^2)^{-b_1/2b_0^2}\,\exp\left\{
   -\fraction{1}{2b_0\gbar^2}\right\}
     \nonumber\\
   &&\times\exp\left\{-\int_0^{\gbar}
   \rmd x \left[\fraction{1}{\beta(x)}
              +\fraction{1}{b_0x^3}
              -\fraction{b_1}{b_0^2x}\right]\right\}, 
   \label{Lambda}\\[1ex]
  M_s &=& \mbar_s\,(2b_0\gbar^2)^{-d_0/2b_0}\exp\left\{-\int_0^{\gbar} 
  \rmd x \left[\fraction{\tau(x)}{\beta(x)}-\fraction{d_0}{b_0x}\right]\right\}.
  \label{RGImass}
\end{eqnarray}
One easily checks that the above expressions for 
$\Lambda$ and $M_s$ 
are indeed solutions of the Callan-Symanzik equation~(\ref{CS}).

\subsection{Finite renormalizations}

Any two mass independent renormalization schemes can be related
by a finite parameter renormalization of the form
\begin{eqnarray}
  \mu'&=&c\mu,\qquad c > 0,\\[1ex]
  \gr'&=&\gr \sqrt{{\cal X}_{\rm g}(\gr)},\\[1ex]
 \mrs'&=&\mrs {\cal X}_{\rm m}(\gr),\quad s=1,\ldots,\Nf, 
\end{eqnarray}
where we assumed the quark mass matrix to be diagonal in both schemes.
The invariance of a physical observable $P$ under such a change
of variables implies the existence of a function $P'$ such that
\begin{equation}
   P'\bigl(\mu'(\mu),\gr'(\gr),\{\mrs'(\gr,\mrs)\}\bigr)
 = P(\mu,\gr,\{\mrs\}),
\end{equation}
and $P'$ satisfies the Callan-Symanzik equation in the primed scheme
with renormalization group functions $\beta'$ and $\tau'$, 
given by
\begin{eqnarray}
  \beta'(\gr')&=& \left\{\beta(\gr)\fraction{\partial\gr'}{\partial\gr}
                  \right\}_{\gr=\gr(\gr')},
  \label{betatransf}\\
  \tau'(\gr') &=& \left\{\tau(\gr)+\beta(\gr)
                  \fraction{\partial}{\partial\gr}\ln{\cal X}_{\rm m}(\gr)
		  \right\}_{\gr=\gr(\gr')}.
  \label{tautransf}
\end{eqnarray}
In perturbation theory, the finite renormalization constants 
${\cal X}_{\rm g}$ and ${\cal X}_{\rm m}$ are expanded according to
\begin{equation}
  {\cal X}(\gr)\buildrel{\gr}\rightarrow0\over\sim
  1+\sum_{k=1}^{\infty}{\cal X}^{(k)}\gr^{2k}.
\end{equation}
Inserting into eqs.~(\ref{betatransf}),(\ref{tautransf}) 
one finds that $b_0$, $b_1$ and $d_0$ 
are the same in both schemes (these are the ``universal" coefficients), 
while all other coefficients are scheme dependent. 
In particular, the two-loop anomalous quark mass
dimensions are related by
\begin{equation}
  d_1'= d_1 +2b_0 {\cal X}_{\rm m}^{(1)}-d_0 {\cal X}_{\rm g}^{(1)}.
  \label{d1pert}
\end{equation}
Renormalization group invariant parameters can also be formed in the 
primed scheme  and one finds
\begin{eqnarray}
  \Lambda' &=& \Lambda\exp\left\{{\cal X}_{\rm g}^{(1)}/2b_0\right\},
  \label{RGItransf1}\\  
  M_s'&=& M_s,\quad s=1,\ldots,\Nf.
  \label{RGItransf2}
\end{eqnarray}
Therefore, given these parameters in some renormalization 
scheme, they are {\em exactly} known in any other scheme by just computing
the one-loop relation between the renormalized coupling constants.
Moreover, any physical observable can be considered a function 
of the renormalization group invariant parameters, i.e.~there 
exists a function $\hat P$ such that
\begin{equation}
  \hat P(\Lambda,\{M_s\}) 
 = P(\mu,\gr,\{\mrs\}).
\end{equation}
It thus appears natural to regard the non-perturbatively defined
renormalization group invariants 
$\{M_s\}_{s=1,\ldots,\Nf}$ and $\Lambda$ (in some renormalization scheme) 
as the fundamental parameters of QCD.

\subsection{Determination of the RGI parameters}

In sect.~3 we will review the SF scheme, 
which is a non-perturbatively defined mass-independent
renormalization scheme. 
The non-perturbative evolution of the running parameters
in this scheme  can be traced using numerical simulations and
the functions $\gbar$ and $\mbar_s$ 
are then known up to some high energy scale $q$. 
One may look for the onset of perturbative evolution, 
and, once arrived in the perturbative regime, 
link the running parameters 
to the renormalization group invariants,
using eqs.~(\ref{Lambda}), (\ref{RGImass}) and the 
perturbative expansion of $\beta$ and $\tau$.
In order to achieve the 
desired precision this requires the knowledge 
of the two-loop anomalous dimension $\d1SF$ 
and the coefficient $\b2SF$ of the 
$\beta$-function in the SF scheme~\cite{RainerEtAl}. 

The easiest method to obtain $\d1SF$ 
consists in calculating the one-loop relation between 
the renormalized parameters in the SF scheme and 
some other scheme in which $d_1$ is already known [cf.~eq.~(\ref{d1pert})]. 
For the purpose of our perturbative study the $\msbar$ scheme
is an appropriate reference scheme and we may thus use the
result~(\ref{d1msbar}).

\section{The SF scheme}

We consider QCD on a finite (Euclidean) space-time manifold of
size $T\times L^3$ with Schr\"odinger functional boundary
conditions for the fields~\cite{LNWW-paperI}. 
All dimensionful quantities are defined in units of $L$. In particular,
for a given correlation function, the ratio $\rat=T/L$ is assumed to be
fixed so
that there is (apart from quark masses) only a single scale, $L$, 
in the theory, which plays the r\^ole of the inverse normalization 
mass, $L=1/\mu$~\cite{letter}. 

In addition to dependence on the geometrical ratio $\rat$, 
correlation functions will also be functions of
the parameter $\theta$ appearing in the definition of the
spatial boundary conditions on the quark fields (see e.g. eq.(4.8)
in ref.~\cite{paperI}). 
One is completely free to employ different 
choices of the parameters $\rat$ and $\theta$ 
for the definitions of different physical quantities.
This is a very convenient aspect of the SF framework
which we will exploit; 
it can however lead to some slight abuse of notation which we hope 
is not confusing in our presentation.

\subsection{Renormalized coupling}

We start with the definition of the renormalized coupling
constant in the SF scheme. This was introduced in ref.~\cite{LNWW} 
for the pure gauge theory. The case of $N=3$ colors, to which we 
restrict attention in this paper, was considered in detail
in ref.~\cite{su3}, and we take over the background gauge field 
and choice of the parameter $\rat=1$ specified there.

These choices were also made for the extension of the
definition of the SF coupling to full QCD  
discussed in ref.~\cite{StefanRainer}. 
There the coupling was renormalized at
the physical value of the quark mass, thus leading to 
a quark mass dependent $\beta$ function. Here we will deviate
from this definition and adopt a mass independent renormalization
scheme as advocated in ref.~\cite{paperI}. 
Renormalizing the coupling at vanishing quark mass 
leads to the quark mass independent 
relation between the renormalized couplings 
\begin{equation}
 \gbar_{\smallSF}^2(L)=\gbar_{\msbar}^2(q){\cal X}_{\rm
g}(\gbar_{\msbar}(q))\,. 
\end{equation}
The SF coupling has been computed to one-loop order of perturbation 
theory in refs.~\cite{su3,StefanRainer},
\begin{equation}
  {\cal X}^{(1)}_{\rm g}=2b_0\ln(qL)-{1\over 4\pi}(c_{1,0}+c_{1,1}\Nf),
  \label{Xg}
\end{equation}
where the coefficient $c_{1,1}$ depends on the parameter $\theta$. 
As a result of the detailed study in 
ref.~\cite{StefanRainer} the particular choice
$\theta=\pi/5$ was recommended for the definition of the
SF coupling in QCD, in which case one obtains 
\begin{equation}
  c_{1,0}=1.25563(4),\qquad  c_{1,1}=0.039863(2).
\end{equation}
In this paper we will also adopt this choice, but
it will be clear where numerical results will change
if in future simulations another choice of $\theta$ is 
employed for $\gbar_{\smallSF}$, and in fact it turns out that 
our general conclusions are not dependent on this.

\subsection{The r\^ole of the PCAC relation}

In QCD with $\Nf\geq 2$ quark flavors the 
PCAC relation provides an attractive starting point for the 
non-perturbative definition of a renormalized quark mass.
Denoting the isospin non-singlet axial current and density
by $A_\mu^a$ and $P^a$, respectively, 
the PCAC relation (for mass degenerate quarks),
\begin{equation}
  \partial_\mu A_\mu^a = 2mP^a,
  \label{PCAC}
\end{equation}
is a local relation between composite fields which is expected 
to hold when inserted in Euclidean correlation functions
up to contact terms.

The normalization of the axial current is conventionally fixed
by requiring current algebra relations to assume their
canonical form~\cite{BochicchioEtAl,paperIV}. 
The non-linearity of these relations
furthermore implies that the anomalous dimension of the axial current 
vanishes and so does the total anomalous dimension 
of the right hand side of eq.~(\ref{PCAC}).
Therefore, a renormalized quark mass can be defined 
through the PCAC relation by providing an independent 
renormalization condition for the axial density.
Below the running quark mass in the SF scheme 
will be defined along these lines by using correlation 
functions derived from the QCD Sch\"odinger functional.

\subsection{The renormalized axial density}

To define the renormalized axial density in the SF scheme we
regularize the theory and
choose the framework of O($a$) improved lattice QCD
as discussed in ref.~\cite{paperI}. 
This choice is motivated by the 
corresponding non-perturbative studies~\cite{RainerEtAl},
for which a perturbative investigation of the cutoff
effects provides complementary  information (cf. sect.~4).
We emphasize, however, that the results for the 
anomalous dimension are independent of this choice, 
and e.g.~dimensional regularization would have been a practical alternative.
In the following we use notations and conventions as
in ref.~\cite{paperI} without further notice.

The renormalized O($a$) improved axial density has
the form
\begin{equation}
  (\pr)^a =\zp(1+\bp a\mq)P^a,\qquad P^a=\psibar \gamma_5\frac12 \tau^a\psi,
\end{equation}
where $\tau^a$ are the Pauli matrices acting in flavor space.
If chosen appropriately the improvement coefficient $\bp$  cancels cutoff
effects in on-shell correlation functions which are 
proportional to the subtracted bare quark mass $\mq=m_0-m_c$.

To define $\zp$ we recall the definition 
of the bare correlation functions $\fp$ and $f_1$,
\begin{eqnarray}
  \fp(x_0)&=&-a^6\sum_{\bf y,z}
  \frac{1}{3}\langle P^a(x)\,
  \zetabar({\bf y})\dirac{5}\frac{1}{2}\tau^a\zeta({\bf z})\rangle,
  \label{fp}\\
  \f1 &=&-{{a^{12}}\over{L^6}}
  \sum_{\bf u,v,y,z}
  \frac{1}{3}
  \langle\zetabarprime({\bf u})
  \dirac{5}\frac{1}{2}\tau^a\zetaprime({\bf v})
  \zetabar({\bf y})\dirac{5}\frac{1}{2}\tau^a\zeta({\bf z})\rangle.
\end{eqnarray}
The boundary source fields $\zeta,\zetabar$  and  $\zeta',\zetabar'$
are renormalized multiplicatively with a common renormalization 
constant $\zzeta$~\cite{StefanII,paperI}.
One may therefore define the renormalization constant of the 
axial density through the ratio~\cite{letter,Martintalk}
\begin{equation}
  \Zp(g_0,L/a)=c{\sqrt{f_1}\over \fp(T/2)},
  \label{Zp}
\end{equation}
at vanishing quark mass $\mq=0$ and for vanishing 
boundary gauge fields. Here, the constant $c$ is chosen such
that $\zp=1$ holds exactly at tree-level of perturbation theory.
Using the notation of refs.~\cite{paperII,paperV} with the
modifications as indicated in the appendix we obtain
\begin{equation}
  c={u_0\over\sqrt{t_0}} = \sqrt{N}+\rmO(a^2).
\end{equation}
In general, $c$ is a computable constant 
for a given lattice size and  values of $\rat$ and $\theta$.

The implicit dependence of $\Zp$ upon $\rat$ and $\theta$
will be discussed later. Here we emphasize that $\Zp$ is 
quark mass independent. Therefore, not only the $\beta$
function but also the anomalous dimension
\begin{equation}
 \tau_{\smallSF}(\gr)=
 \left.L\fraction{\partial\ln\zp(g_0,L/a)}{\partial L}
 \right\vert_{g_0=g_0(\gr)}
 \label{tauSF}
\end{equation}
is quark mass independent. Here $\gr$ stands for $\gbar_{\smallSF}(L)$,
and its (mass independent) relation to $g_0$ is currently known to 
one-loop order~\cite{StefanRainer} and to two-loop order in 
quenched QCD~\cite{Bode}. 

In bare perturbation theory the renormalization
constant $\zp$ has an expansion
\begin{equation}
  \zp(g_0,L/a)=1+\sum_{k=1}^\infty \zp^{(k)}(L/a)\, g_0^{2k},
\end{equation}
where in the limit $a/L\to 0$   
the coefficients $\zp^{(k)}$ are polynomials in
$\ln(L/a)$ of degree $k$ up to corrections of O($a/L$).
In particular the coefficient of the logarithmic divergence in 
$\zp^{(1)}$ is given by
the one-loop anomalous quark mass dimension $d_0$, 
and thus we parametrize $\zp^{(1)}$ as 
\begin{equation}
  \zp^{(1)} = \cf z_p(\theta,\rat)-d_0\ln(L/a)+{\rm O}(a/L)\,. 
\label{zp1}
\end{equation} 
Here we have made explicit the dependence of 
the cutoff independent term $z_p$ on 
$\theta$ and $\rat$, which is inherited by any renormalized
quantity involving the pseudoscalar density. 
The Feynman diagrams
contributing to the correlation functions $\fp, f_1$ 
(and hence to $\zp$) at one loop 
order have been discussed in detail in refs.~\cite{paperII,paperV}.
Here we use these results to extract $\zp^{(1)}$ for various
different choices of the parameters $\rat=T/L$ and $\theta$.  
Some typical results for $z_p$ are collected in table~1. 
In section 4 we will also consider
the $\theta, \rat$ dependence of the remaining cutoff terms
in (\ref{zp1}). 
  \begin{table}[t]
%[htb]
\centering
\begin{tabular} {|c|c|c|c|c|c|}
\hline
  &&&&&\\[-1ex]
  $\rat$ & $\theta$& $z_p(\theta,\rat)$ & $z_p|_{{\rm mom}=0}$& $\Delta
z_p$ 
  &$\d1SF/d_0|_{N=3}$ \\[1ex]
\hline
  &&&&&\\[-1.5ex]
  $2$ & $0.0$ & $-0.33879(1)$ & $-0.250$ & $-0.089$ & 
                            $\phantom{-}0.9290-0.0441\Nf$ \\[0.5ex]
%%%%                            $\phantom{-}0.9290-0.0455\Nf$ \\[0.5ex]
%
  $2$ & $0.5$ & $\phantom{-}0.07494(1)$ & $\phantom{-}0.165$ & $-0.090$ & 
                            $-0.5880+0.0478\Nf$ \\[0.5ex]
%%%%                            $-0.5880+0.0464\Nf$ \\[0.5ex]
%
  $1$ & $0.0$ & $-0.11953(2)$ & $-0.031$ & $-0.089 $ &
                       $\phantom{-}0.1251+0.0046\Nf$ \\[0.5ex]
%%%%                       $\phantom{-}0.1251+0.0032\Nf$ \\[0.5ex]
%
  $1$ & $0.5$ & $-0.09281(2)$ & $-0.003$ & $-0.090$ &
                       $\phantom{-}0.0271+0.0105\Nf$ \\[0.5ex]
%%%%                       $\phantom{-}0.0271+0.0091\Nf$ \\[0.5ex]
%
\hline
\end{tabular}
\caption{\footnotesize The finite part of the one-loop renormalization
constant of the axial density, the contribution of zero momentum gluon
exchange, and the corresponding two-loop anomalous dimension.}
\end{table}

\subsection{Renormalized quark masses}

With the O($a$) improved bare axial current
\begin{equation}
    (A_{\rm I})_\mu^a= A^a_{\mu}+\ca a\frac12
                 (\partial^*_{\mu}+\partial_{\mu})P^a,
    \qquad A^a_\mu=\psibar\gamma_\mu\gamma_5\frac12 \tau^a \psi,
   \label{Aimp}
\end{equation}
the renormalized current takes the form,
\begin{equation}
  (\ar)_\mu^a = \za(1+\ba a\mq)(A_{\rm I})_\mu^a.
\end{equation}
To define the renormalized quark mass we first
introduce a bare current quark mass through the PCAC relation
between the unrenormalized fields,
\begin{equation}
  m = \left[{\frac{1}{2}(\drvstar{0}+\drv{0})
      f_{{\rm A}_{\rm I}}(x_0)
      \over 2\fp(x_0)}\right]_{x_0=T/2}.
\end{equation}
Here, the correlation function of the axial current is defined 
as $\fp$ in eq.~(\ref{fp}), but with the axial density 
replaced by the zero component of the (improved) axial current~(\ref{Aimp}).

The renormalized mass in the SF scheme
is defined  through the PCAC relation involving the renormalized 
O($a$) improved fields.
This leads to the relation
\begin{equation}
  \mbar_{\smallSF}(L)= m \fraction{(1+\ba a\mq)\za}{(1+\bp a\mq)\zp}
                     =m\za/\zp +{\rm O}(a).
\end{equation}
In order to compute the 
two-loop anomalous dimension in this scheme we first relate
$\mbar_{\smallSF}(L)$ to the running mass in the $\msbar$ scheme.
To this end we start by combining
the result for the axial current renormalization
constant~\cite{GabrielliEtAl,paperIV}
\begin{equation}
  \za^{(1)}=-0.087344(1)\times\cf, 
\end{equation}
with the one-loop value of the ratio $\mq/m$, which
we obtained in the course of calculations done in ref.~\cite{paperV}.
We then arrive at  
\begin{equation}
  \mbar_{\smallSF}(L)=\mq\big\{1+g_0^2
  \bigl[d_0\ln(L/a)-\bigl(z_p+0.019458(1)\bigr)\cf\bigr]
  +\rmO(g_0^4)\bigr\}.
  \label{mSFpert}
\end{equation}
The corresponding relation for the renormalized $\msbar$ mass
has first been obtained in ref.~\cite{GabrielliEtAl} and
since then verified by many others, including one of the
present authors\footnote{\footnotesize Ref.~\cite{GabrielliEtAl},
although analytically correct, contains a small error in the
quoted numerical results, which is caused by
setting $F_{0001}=1.41$  rather than $F_{0001}=1.310962...$}.
Then using the result
\begin{equation}
  \mbar_{\msbar}(q)=\mq\bigl\{1+g_0^2 
  \bigl[-d_0\ln(aq)+0.122282(1)\times\cf\bigr]+\rmO(g_0^4)\bigr\},
\end{equation}
we obtain the one-loop coefficient
\begin{equation}
  {\cal X}^{(1)}_{\rm m}=d_0\ln(qL)-
  \bigl( z_p+0.141740(2)\bigr)\cf. 
  \label{Xm}
\end{equation}
To finally obtain the two-loop anomalous dimension
$\d1SF$ we may now use the known result for $d_1$
in the $\msbar$ scheme [cf.~eq.~(\ref{d1msbar})] and combine it with
the one-loop coefficients~(\ref{Xg}) and (\ref{Xm}) according
to eq.~(\ref{d1pert}). Proceeding in this way, we have avoided to
expand eq.~(\ref{Zp}) to order $\gr^4$ which would have
required a two-loop computation.
A few numerical values are given in table~1.

\subsection{Zero-momentum gluon exchange}

The results for $\zp^{(1)}$ and $\d1SF$ with  $\rat=2$
show a strong dependence on the parameter 
$\theta$. In the corresponding non-perturbative 
study a similar behavior is only seen 
at very small couplings~\cite{RainerEtAl}. While this
is not a problem in principle, it makes it more difficult to
connect the perturbative regime to low energy physics
along the lines of ref.~\cite{letter}.

A closer look into the one-loop computation reveals that
this strong $\theta$-dependence is almost entirely due to the
exchange of gluons with zero spatial momentum~\cite{Martin}.
These contributions are gauge invariant by themselves and
it is not too difficult to compute them analytically.
Setting 
\begin{equation}
    w=\sqrt{3}\theta\rat,
\end{equation}
we find
\begin{eqnarray}
  z_p(\theta,\rat)\vert_{\rm mom=0} 
  &=&{\rat^3\over192w^4\cosh^2w}\Bigl\{72+24w^2-5w^4-96\cosh{w}
  \nonumber\\[1ex]
  &&\mbox{}+(3w^4+24)\cosh2w-12w\sinh2w\Bigr\}, 
\end{eqnarray}
with the special case
\begin{equation}
  z_p(0,\rat)\vert_{\rm mom=0} = -\frac1{32}\rat^3.
\end{equation}
One notices the overall factor $\rat^3$ which enhances the 
$\theta$-dependence of $z_p$ for large $\rat$ and results in the
strong $\theta$-dependence of $\d1SF$ for $\rat=2$. 
If a subtracted constant is defined through
\begin{equation}
  \Delta z_p = z_p-z_p\vert_{\rm mom=0},
\end{equation}
the remaining $\theta$-dependence is indeed very weak.
We thus recommend to define the renormalized axial 
density with $\rat=1$. Having fixed this parameter
we furthermore choose $\theta=0.5$ which leads to
a conveniently small value of $\d1SF/d_0$ (cf.~table~1).
This completes the definition of the running parameters
in the SF scheme.

\section{One-loop cutoff effects in the step scaling function}

An infinitesimal variation of $\Zp$ with the scale $L$ 
defines the anomalous dimension $\tau_{\smallSF}(\gr)$ [cf.~eq.~(\ref{tauSF})].
In the context of numerical simulations 
it is more convenient to consider finite variations 
of the scale, e.g.~a  change from $L$ to $sL$, with a scale factor $s$.
This leads to the definition of the step scaling function~\cite{letter}
\begin{equation}
  \Sigma_{\rm P}(s,\gr^2,a/L)=
  \left.{\Zp(g_0,sL/a)\over\Zp(g_0,L/a)}\right\vert_{g_0=g_0(\gr)},
\end{equation}
with continuum limit 
\begin{equation}
  \lim_{a\rightarrow 0} \Sigma_{\rm P}(s,\gr^2,a/L) 
   = \sigma_{\rm P}(s,\gr^2),
\end{equation}
to be taken at fixed $\gr=\gbar_{\smallSF}(L)$.
%This, together with the requirement of vanishing quark mass
%defines a renormalization group trajectory, and one may then
%ask how large the cutoff effects are for various observables,
%in particular the step scaling function itself.
%One may then ask how rapidly the limit is approached.

We set the scale factor to $s=2$ in the following.
To one-loop order of perturbation theory we have
\begin{equation}
  \Sigma_{\rm P}(2,\gr^2,a/L)=1+k(L/a)\gr^2+\rmO(\gr^4)\,,
\end{equation}
with
\begin{equation}
  k(L/a)= \Zp^{(1)}(2L/a)-\Zp^{(1)}(L/a).
\end{equation}
Using the notation of refs.~\cite{paperII,paperV} for the
renormalized correlation functions $\fp$ and $ f_1$, 
the one-loop coefficient
on a finite lattice takes the form
\begin{equation}
  \Zp^{(1)}(L/a)=\fraction{t_1}{2t_0}-\fraction{u_1}{u_0}
    +\ctildet^{(1)}\left[\fraction{t_2}{2t_0}-\fraction{u_2}{u_0}\right]
        +a\mc^{(1)}\left[\fraction{t_3}{2t_0}-\fraction{u_3}{u_0}\right].
  \label{Zp_finitelattice}
\end{equation}
In order to see how fast the continuum limit,  
\begin{equation}       
   k(\infty)=-d_0\ln(2),
\end{equation}
is approached we define
\begin{equation}
  \delta_{\rm k}(L/a)=k(L/a)/k(\infty)-1.
\end{equation}
This quantity contains all lattice artifacts at O($\gr^2$).
In the framework of O($a$) improved lattice QCD these
are expected to decrease asymptotically 
with a rate proportional to $a^2/L^2$. 
In the present case, O($a$) improvement is achieved 
by setting $\csw^{(0)}=1$. In particular,
the boundary counterterms proportional to $\ct$ and $\ctildet$
are not needed at this order of perturbation theory, 
owing, in the case of $\ctildet$, to the identity
\begin{equation}
   \fraction{t_2}{2t_0}-\fraction{u_2}{u_0}=0.
   \label{identity}
\end{equation}
Some results for $\delta_{\rm k}$ are tabulated in table~2.
The numerical values have been obtained by inserting  
the exact expressions for the coefficients 
$u_0,u_1,u_3$ and $t_0,t_1,t_3$ 
for the given lattice size,
and the coefficient
$a\mc^{(1)}=-0.2025565(1)\times\cf$ into eq.~(\ref{Zp_finitelattice}).

\begin{table}[t]
%[htb]
\centering
\begin{tabular} {|c|c|c||c|c|}
\hline
  &&&&\\[-1ex]

  $L/a$  &  $\delta_{\rm k}|_{\theta=0}$  
          &  $\Delta\delta_{\rm k}|_{\theta=0}$  
          &  $\delta_{\rm k}|_{\theta=0.5}$ 
          &  $\Delta\delta_{\rm k}|_{\theta=0.5}$ \\[0.5ex] 
\hline
\multicolumn{5}{|c|}{$\rat=2$} \\
\hline
 $4$   &  $-0.3084$  &             $-0.0025$  
       &  $-0.2456$  &             $-0.0702$ \\
 $6$   &  $-0.2292$  &             $-0.0067$  
       &  $-0.1499$  &             $-0.0444$ \\
 $8$   &  $-0.1449$  &  $\hphantom{+}0.0046$
       &  $-0.0893$  &             $-0.0209$ \\
$10$   &  $-0.0974$  &  $\hphantom{+}0.0076$ 
       &  $-0.0584$  &             $-0.0109$ \\
$12$   &  $-0.0696$  &  $\hphantom{+}0.0076$ 
       &  $-0.0412$  &             $-0.0064$ \\
$14$   &  $-0.0522$  &  $\hphantom{+}0.0068$ 
       &  $-0.0307$  &             $-0.0042$ \\
$16$   &  $-0.0405$  &  $\hphantom{+}0.0059$ 
       &  $-0.0239$  &             $-0.0030$ \\
%$18$   &  $-0.0324$  &  $\hphantom{+}0.0051$ 
%       &  $-0.0191$  &             $-0.0022$ \\
%$20$   &  $-0.0265$  &  $\hphantom{+}0.0044$ 
%       &  $-0.0157$  &             $-0.0018$ \\
%$22$   &  $-0.0221$  &  $\hphantom{+}0.0039$ 
%       &  $-0.0131$  &             $-0.0015$ \\
%$24$   &  $-0.0186$  &  $\hphantom{+}0.0034$ 
%       &  $-0.0111$  &             $-0.0012$ \\
%
\hline\hline
\multicolumn{5}{|c|}{$\rat=1$} \\
\hline
 $4$ &  $\phantom{-}0.2040$ &  $\phantom{-}0.0650$ 
     &  $\phantom{-}0.2136$ &  $\phantom{-}0.0140$ \\
 $6$ &            $-0.0121$ &  $\phantom{-}0.0126$ 
     &  $\phantom{-}0.0208$ &            $-0.0198$ \\
 $8$ &            $-0.0253$ &  $\phantom{-}0.0129$ 
     &            $-0.0026$ &            $-0.0102$ \\
$10$ &            $-0.0215$ &  $\phantom{-}0.0123$ 
     &            $-0.0062$ &            $-0.0049$ \\
$12$ &            $-0.0171$ &  $\phantom{-}0.0107$ 
     &            $-0.0064$ &            $-0.0025$ \\
$14$ &            $-0.0137$ &  $\phantom{-}0.0090$ 
     &            $-0.0058$ &            $-0.0014$ \\
$16$ &            $-0.0111$ &  $\phantom{-}0.0076$ 
     &            $-0.0052$ &            $-0.0009$ \\

\hline
\end{tabular}
\caption{\footnotesize The one-loop cutoff effects $\delta_{\rm k}$
 in the step scaling function,  
 with and without zero momentum gluon contributions, for the same
 choice of parameters as in sect.~3.} 
\end{table}

Lattice artifacts appear to be reasonably small for all parameter
choices considered. However, we observe that cutoff effects for 
$\rat=2$ are generally larger than for $\rat=1$. 
Furthermore the asymptotic O($a^2$)
decay seems to set in earlier for $\rat=2$.
It turns out that both effects are largely due
to the zero spatial momentum gluon exchange contributions.
Subtracting these contributions from  $\delta_{\rm k}$ we define
\begin{equation}
  \Delta\delta_{\rm k}\defeq [k(L/a)-k(L/a)\vert_{\rm mom=0}]/k(\infty)-1.
\end{equation}
and list the numerical values in table~2.
The subtraction terms are computed numerically for the given lattice
size. In the special case $\theta=0$ we also obtained a
compact analytical formula,
\begin{equation}
   k(L/a)\bigl\vert_{{\rm mom=0},\,\theta=0}=
  \left(\fraction{3\rat a^2}{16L^2}-\fraction{63a^3}{64L^3}\right)\cf.
  \label{zero}
\end{equation}
For $\rat=2$ one clearly sees that the cutoff effects 
are dominated by the zero-momentum contributions.
In the case $\theta=0$ it is the explicit factor $\rat$ 
in eq.~(\ref{zero}) which enhances the cutoff effects and 
also explains the early onset of the O($a^2$) behavior for $\rat=2$.
We conclude by noting that cutoff effects with the 
parameters $\rat=1$ and $\theta=0.5$ are indeed quite small, 
a fact that partially motivated this choice.

\section{Summary}

We have carried out a perturbative investigation of 
a two-parameter family of running quark masses in the SF scheme. 
Its definition is based on the PCAC relation between correlation functions
derived from the Schr\"odinger functional, together with an independent
renormalization condition for the axial density.
At one-loop order of perturbation theory  and for asymmetric 
space-time volumes with $T=2L$, many correlation functions 
show a strong dependence on the parameter $\theta$.
Its origin could be traced back to the contribution of gluon exchange with
vanishing spatial momentum. Non-perturbatively this behavior
is only matched at very short distances~\cite{RainerEtAl}, making 
it more difficult to apply the strategy of non-perturbative
renormalization as outlined in refs.~\cite{letter,Martintalk}.
However, setting $T=L$ completely eliminates this problem 
and also leads to a reasonably small two-loop anomalous dimension $\d1SF$. 
We believe that this situation is generic and thus generally recommend the 
choice of symmetric space-time volumes for the study of 
scale dependent renormalization constants in the SF scheme.

In the present case we made the additional choice
of $\theta=0.5$ thus completing the definition of the 
running quark mass. One-loop cutoff 
effects in the step-scaling function 
are found to be reasonably small if O($a$) improved 
lattice QCD is used as a regularization.
Finally, we mention that a corresponding 
non-perturbative study of the running quark
mass is in progress~\cite{RainerEtAl}, and preliminary results
have been reported in ref.~\cite{Martintalk}.

%While the
%numerical simulations are currently done in the quenched approximation,
%an extension to a fixed non-zero number of quark flavors does
%not introduce any new conceptual problems. There is, however, a drawback
%of using a quark mass independent renormalization scheme.
%While renormalization group considerations are simplified, the
%decoupling of heavy quark flavors has to be built in by hand, 
%by imposing appropriate matching conditions at heavy quark thresholds. 
%While there exist standard procedures 
%in perturbation theory~\cite{BernreutherWetzel},
%a practical non-perturbative strategy has not yet been devised.

\vskip 1.5ex

This work is part of the ALPHA collaboration research program.
We thank Martin L\"uscher, Rainer Sommer and Hartmut Wittig 
for helpful discussions and a critical reading of the manuscript. 
S. Sint acknowledges support by the U.S. Department of Energy
(contracts DE-FG05-85ER250000 and DE-FG05-96ER40979).

\pagebreak

\appendix
\renewcommand{\thesection}{Appendix~A}
\section{}
\renewcommand{\thesection}{A}

The expansions of the functions $t_i,u_i,v_i,w_i,y_i \quad i\ne 1$
up to corrections of order $(a/L)^2$ for general $\rat=T/L$
are as given in appendix B 
of ref.~\cite{paperII} and in appendix A of ref.~\cite{paperV}
except that $\si,\co$ are now defined by
\begin{eqnarray}
  \si&=&\sinh(\sqrt{3}\,\theta\rat),\\[2ex]
  \co&=&\cosh(\sqrt{3}\,\theta\rat),
\end{eqnarray}
and the expressions for $i=3$ now read
\begin{eqnarray}
  t_3&=&-{2N\si\over\sqrt{3}\theta\co^3}{L\over a}
  +{2N\over\co^2}\nonumber\\[2ex]
  &&\mbox{}-\Bigl\{ {19N\theta\si\over3\sqrt{3}\co^3}
  -{4\rat N\theta^2(1-2\si^2)\over3\co^4}\Bigr\}{a\over L},\\[2ex]
  u_3&=&-{N\si\over\sqrt{3}\theta\co^2}{L\over a}
  -\Bigl\{ {N\theta\si\over 6\sqrt{3}\co^2} 
  -{2\rat N\theta^2\over 3\co^3}(1-\si^2)\Bigr\}{a\over L},\\[2ex]
  v_3&=&-{N\si(\co-2)\over\sqrt{3}\theta\co^3}{L\over a}
  -\Bigl\{ {N\theta\si(\co-2)\over 6\sqrt{3}\co^3}\nonumber\\[2ex]
  &&\mbox{}+{2\rat N\theta^2\over 3\co^4}
      (\co^3-4\co^2-2\co+6)\Bigr\}{a\over L},\\[2ex]
  w_3&=&{2N\over\co}{L\over a}
  +\Bigl\{ {3N\theta^2\over \co}
  +{4\rat N\theta^3\si\over \sqrt{3}\co^2}\Bigr\}{a\over L},\\[2ex]
  y_3&=&-{N\si(\co+2)\over3\sqrt{3}\theta\co^3}{L\over a}
  -\Bigl\{ {N\theta\si(\co+2)\over 18\sqrt{3}\co^3}\nonumber\\[2ex]
  &&\mbox{}+{2\rat N\theta^2\over
9\co^4}(\co^3+4\co^2-2\co-6)\Bigr\}{a\over L}.
\end{eqnarray}

\vfill
\eject

% List of references

\end{document}